\documentstyle[graphicx,12pt]{article}
\begin{document}

\small
\hoffset=-1truecm
\voffset=-2truecm
\title{\bf On the Casimir effect for parallel plates in the spacetime
with one extra compactified dimension}
\author{Hongbo Cheng\footnote {E-mail address:
hbcheng@public4.sta.net.cn}\\
Department of Physics, East China University of Science and
Technology,\\ Shanghai 200237, China}

\date{}
\maketitle

\begin{abstract}
In this paper, the Casimir effect for parallel plates in the
presence of one compactified universal extra dimension is
reexamined in detail. Having regularized the expressions of
Casimir force, we show that the nature of Casimir force is
repulsive if the distance between the plates is large enough,
which is disagree with the experimental phenomena.
\end{abstract}
\vspace{8cm} \hspace{1cm}
PACS number(s): 11.10.Kk, 04.62.+v

\newpage

The higher dimensional spacetime has become a powerful ingredient
for unifying the interactions. Kaluza and Klein put forward the
idea that our universe has more than four dimensions about 80
years ago [1, 2]. Their theory introduced an additional
compactified dimension in order to unify gravity and classical
electrodynamics. Now the string theory is developed to unify the
quantum mechanics and gravity with the help of introducing seven
extra spatial dimensions. It is interesting that the order of the
compactification scale of the extra dimensions has not been
confirmed. Some branches of string theory expect that the radii of
the compactified universal extra dimensions should be Planck size
which is beyond our experimental reach of today and near future
[3, 4]. In some approaches large extra dimensions were also
invoked for providing a breakthrough of hierarchy problem [5-10].
The gauge fields may be localized on a four-dimensional brane, our
real universe, and only gravitons can propagate in the extra space
transverse to the brane [8, 9]. It is possible to probe the large
additional spatial dimensions.

The Casimir effect is a fundamental aspect of quantum field theory
in confined geometries and the physical manifestation of
zero-point energy [11-21]. The precision of the measurement has
been greatly improved experimentally [22-25]. Therefore the
Casimir effect can become a useful method for the study of a lot
of topics. The magnitude of cosmological constant can be estimated
with Casimir effect [26, 27]. The effect was also applied in the
context of string theory [28-31]. Recently more progresses of the
effect were made to investigate the properties of the spacetimes
with extra dimensions [32-33]. As the first step of
generalization, some topics in five-dimensional spacetimes were
studied and the useful results were obtained [5-9]. Probing the
possible existence and size of the extra dimension by means of
Casimir effect attracts more attentions of the physical community
[32-33]. The expressions of Casimir force between two parallel
plates in the presence of one extra dimension differ from the
force in the case without additional spatial dimensions. By
comparison to experimental data the size of the universal extra
dimension can be restricted to $L\leq10nm$ for one extra dimension
only when the distance between plates is very small [33].

The Casimir effect in the presence of a compactified universal
extra dimension needs to be explored in detail. In this paper we
reexamine the Casimir effect for parallel plates in the universe
with only one additional dimension carefully. Having regularized
the total energy, we obtain the Casimir energy, and then Casimir
force. We find that the Casimir force is similar to the
experimental data when the plates approach very close, and the
upper limit of $L\leq10nm$ on the extension of additional spatial
dimension can also be obtained. According to our studies, the
expression for Casimir force in the case with one extra dimension
also shows that the plates will repulse each other for their large
enough gap, but in the experiments the repulsive force has not
appeared within the region of distance between plates [22-25]. The
five-dimensional spacetime could not be feasible, and just be
thought as a toy model. Here we rederive the Casimir energy and
Casimir force with one extra dimension. We discuss these
expressions for various ratio of plates distance and the radius of
extra dimension. Finally the conclusions are emphasized.

In the Kaluza-Klein (KK) approach we study the scalar field in the
system consisting of two parallel plates in the spacetime with
only one extra dimension. Along the extra dimension the wave
vectors of the field have the form $k_{n}=\frac{n}{L}$ with $n$
and $L$ being an integer and the radius of the extra dimension
respectively. At the plates the fields satisfy the Dirichlet
condition, leading the wave vector in the directions restricted by
the plates to be $k_{l}=\frac{\pi l}{R}$, $l$ a positive integer
and $R$ the separation of the plates. Under the conditions
mentioned above, the zero-point fluctuations of the fields can
give rise to observable Casimir forces.

In the case of one extra dimension, the total energy density of
the fields in the interior of system is thus given by

\begin{equation}
\varepsilon=\int\frac{d^{2}k}{(2\pi)^{2}}\sum_{l=1}^{\infty}
\sum_{n=0}^{\infty}\frac{1}{2}\sqrt{k^{2}+\frac{l^{2}\pi^{2}}{R^{2}}
+\frac{n^{2}}{L^{2}}}
\end{equation}

\noindent where

\begin{equation}
k^{2}=k_{1}^{2}+k_{2}^{2}
\end{equation}

\noindent $k_{1}$ and $k_{2}$ are the wave vectors in directions
of the unbound space coordinates. Following [12-15, 19], Eq.(1)
becomes,

\begin{equation}
\varepsilon=-\frac{1}{8\pi^{\frac{3}{2}}}\Gamma(-\frac{3}{2})
E_{2}(\frac{\pi^{2}}{R^{2}},\frac{1}{L^{2}};-\frac{3}{2})
-\frac{\pi^{\frac{3}{2}}}{8}\Gamma(-\frac{3}{2})\zeta(-3)\frac{1}{R^{3}}
\end{equation}

\noindent where Epstein zeta function
$E_{p}(a_{1},a_{2},\cdot\cdot\cdot,a_{p};s)$ is defined as,

\begin{equation}
E_{p}(a_{1},a_{2},\cdot\cdot\cdot,a_{p};s)
=\sum_{\{n\}=1}^{\infty}(\sum_{j=1}^{p}a_{j}n_{j}^{2})^{-s}
\end{equation}

\noindent here $\{n\}$ represents a short notation of $n_{1}$,
$n_{2}$, $\cdot\cdot\cdot$, $n_{p}$, $n_{\alpha}$ a positive
integer.

By regularizing Eq.(3), we obtain the Casimir energy density,

\begin{eqnarray}
\varepsilon_{C}=[\frac{1}{16\pi^{5}}\Gamma(2)\zeta(4)
-\frac{1}{16\pi^{\frac{13}{2}}}\mu\Gamma(\frac{5}{2})\zeta(5)\hspace{2cm}
\nonumber\\-\frac{1}{4\pi^{2}\mu}\sum_{n_{1},n_{2}=1}^{\infty}
(\frac{n_{2}}{n_{1}})^{2}K_{2}(2\mu n_{1}n_{2})
-\frac{\pi^{2}}{720}\frac{1}{\mu^{3}}]\frac{1}{L^{3}}
\end{eqnarray}

\noindent where

\begin{equation}
\mu=\frac{R}{L}
\end{equation}

\noindent and $K_{\nu}(z)$ is the modified Bessel functions of the
second kind and falls exponentially with $z$. The terms with
series converge very quickly and only the first several summands
need to be taken into account for numerical calculation to further
discussions. Having derived in detail, we are able to prove that
the Eq. (5) can become the expression of Casimir energy for a
massive scalar field with mass $m=\frac{n}{L}$ [17]. We analyze
the Casimir energy (5) in the limits. If the plates distance is
much larger or less than the radius of universal extra dimension,
the expression for the Casimir energy becomes,

\begin{equation}
\varepsilon_{C}(\mu\gg1)=-\frac{1}{16\pi^{\frac{13}{2}}}
\mu\Gamma(\frac{5}{2})\zeta(5)\frac{1}{L^{3}}
\end{equation}

\noindent or

\begin{equation}
\varepsilon_{C}(\mu\ll1)=-\frac{\pi^{2}}{720}\frac{1}{\mu^{3}}\frac{1}{L^{3}}
\end{equation}

\noindent respectively. In the case of $\mu\ll1$, the first more
summands in Eq. (5) need to be considered, but the value of the
third term is much smaller than that of the last term because the
term with series also converge very quickly according to the
property of the modified Bessel functions of the second kind
$K_{\nu}(z)$. The numerical calculations of the Casimir energy (5)
lead to the data presented in Figure 1. The figure shows that the
sign of the Casimir energy depending on the ratio $\mu$ keeps
negative no matter what the value the ratio $\mu$ is.

It is certainly fundamental to discuss the Casimir force between
the plates in the background with one extra dimension in order to
compare our results with the experimental phenomenon. The Casimir
force is denoted as
$f_{C}=-\frac{1}{L}\frac{\partial\varepsilon_{C}}{\partial\mu}$,
so its expression is obtained as follow,

\begin{eqnarray}
f_{C}=\{\frac{1}{16\pi\pi^{\frac{13}{2}}}\Gamma(\frac{5}{2})\zeta(5)
-\frac{1}{4\pi^{2}\mu^{2}}\sum_{n_{1},n_{2}=1}^{\infty}
(\frac{n_{2}}{n_{1}})^{2}K_{2}(2\mu n_{1}n_{2})\hspace{1.5cm}\nonumber\\
-\frac{1}{4\pi^{2}\mu}\sum_{n_{1},n_{2}=1}^{\infty}\frac{n_{2}^{3}}{n_{1}}
[K_{1}(2\mu n_{1}n_{2})+K_{3}(2\mu
n_{1}n_{2})]-\frac{\pi^{2}}{240\mu^{4}}\}\frac{1}{L^{4}}
\end{eqnarray}

\noindent Of course the equation (9) needs to be analyzed in the
limits. According to the property of function $K_{\nu}(z)$ and
discussion above, the expression for the Casimir force becomes,

\begin{equation}
f_{C}(\mu\ll1)=-\frac{\pi^{2}}{240\mu^{4}}\frac{1}{L^{4}}
\end{equation}

\noindent for smaller separation. In the case of larger ones,

\begin{equation}
f_{C}(\mu\gg1)=\frac{1}{16\pi^{\frac{13}{2}}}
\Gamma(\frac{5}{2})\zeta(5)\frac{1}{L^{4}}>0
\end{equation}

\noindent which means that there exists a repulsive force in the
system if the plates separation is large enough. According to (9)
the Casimir force for parallel plates in the spacetime with one
extra dimension is depicted in Figure 2. After proceeding the
numerical calculation, we obtain the special ratio $\mu_{f}=5.343$
satisfying $f_{C}(\mu=\mu_{f})=0$. When $\mu<\mu_{f}$, then
$f_{C}<0$, and in the opposite $f_{C}>0$ for $\mu>\mu_{f}$. In the
case of $\mu\ll\mu_{f}$, the part of curve of Casimir force for
two parallel plates depicted in Figure 2 can be used to estimate
the size of the extra dimension by comparison to the experimental
data. K. Poppenhaeger et al have carried out the study for the
case that the plates locate very close each other [33]. They
compared the curves for different value of size of extra dimension
$L$ with real experimental data to discover that good agreement
with the data can only be obtained if the upper limit on the
radius of the only one additional dimension is $L\leq10nm$. In the
spacetime without additional spatial dimensions, the standard
Casimir force between parallel plates is
$f_{C0}=-\frac{\pi^{2}}{240}\frac{1}{R^{4}}$. The ratio of the
Casimir force (with one extra dimension) to the standard ones
(without extra dimensions) is $q=\frac{f_{C}}{f_{C0}}$ and depends
on $\mu$. The curve of ratio $q$ is shaped in Figure 3 and shows
the extra-dimension correction to the standard Casimir force. The
larger the plates separation is, the greater the correction is.
However we should not neglect that the repulsive Casimir force
denoted as $f_{C}>0$ under $\mu>\mu_{f}$ is excluded in the
practice [22-25]. We must point out that the experiment is always
performed on electromagnetic fields that may obey more complicated
boundary conditions than the scalar field we consider here,
according to the 5-dimensional Kaluza-Klein theory the expressions
of Casimir force for different kinds of fields with different
boundary conditions will certainly be different, and the special
ratio $\mu_{f}$ will also not be equal to what is obtained above,
but the repulsive Casimir force must appear when the separation of
two plates is sufficiently large. Our conclusion that there must
exist the repulsive Casimir force in the Universe with only one
extra dimension is kept.

In conclusion, the model that the spacetime with only one extra
dimension can not be realistic. Having studied the Casimir effect
for two parallel plates in the universe with one additional
dimension, we find analytically that there must exist the
repulsive Casimir force between the plates when the separation is
large enough. The experimental data show that no repulsive Casimir
force appeared. Therefore the results obtained form the
Kaluza-Klein theory including only one compactified universal
extra dimension disagree with the experimental results. The
related topics need further research.

\vspace{3cm}

\noindent\textbf{Acknowledgement}

The author thank Professor I. Brevik and Professor E. Elizalde for
helpful discussions. This work is supported by the Basic Theory
Research Fund of East China University of Science and Technology,
grant No. YK0127312 and the Shanghai Municipal Science and
Technology Commission No. 04dz05905.

\newpage
\begin{figure}
\setlength{\belowcaptionskip}{10pt} \centering
  \includegraphics[width=15cm]{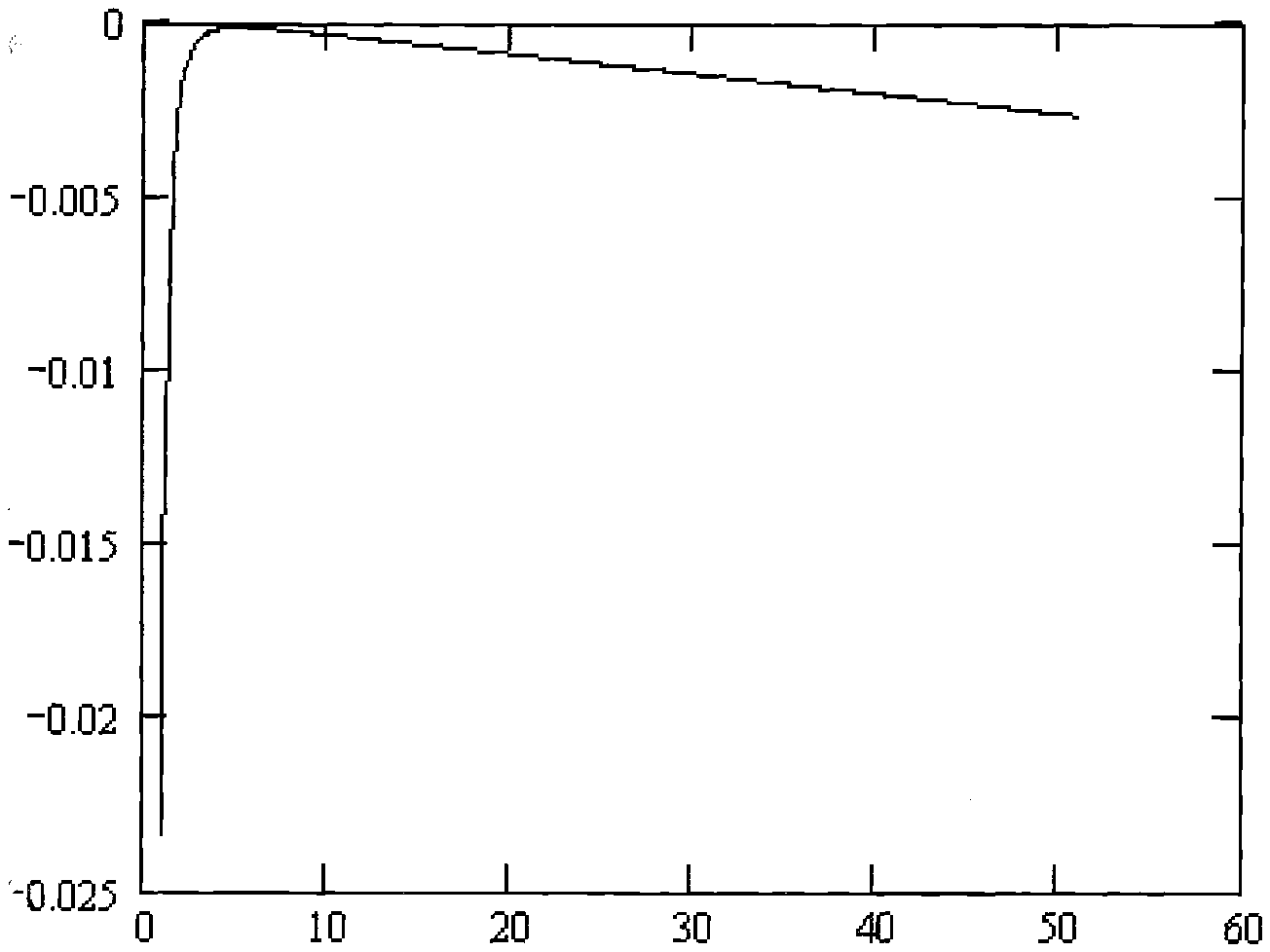}
  \caption{The Casimir energy density versus $\frac{R}{L}$
  for two parallel plates in the spacetime with one compactified extra dimension.}
\end{figure}

\newpage
\begin{figure}
\setlength{\belowcaptionskip}{10pt} \centering
  \includegraphics[width=15cm]{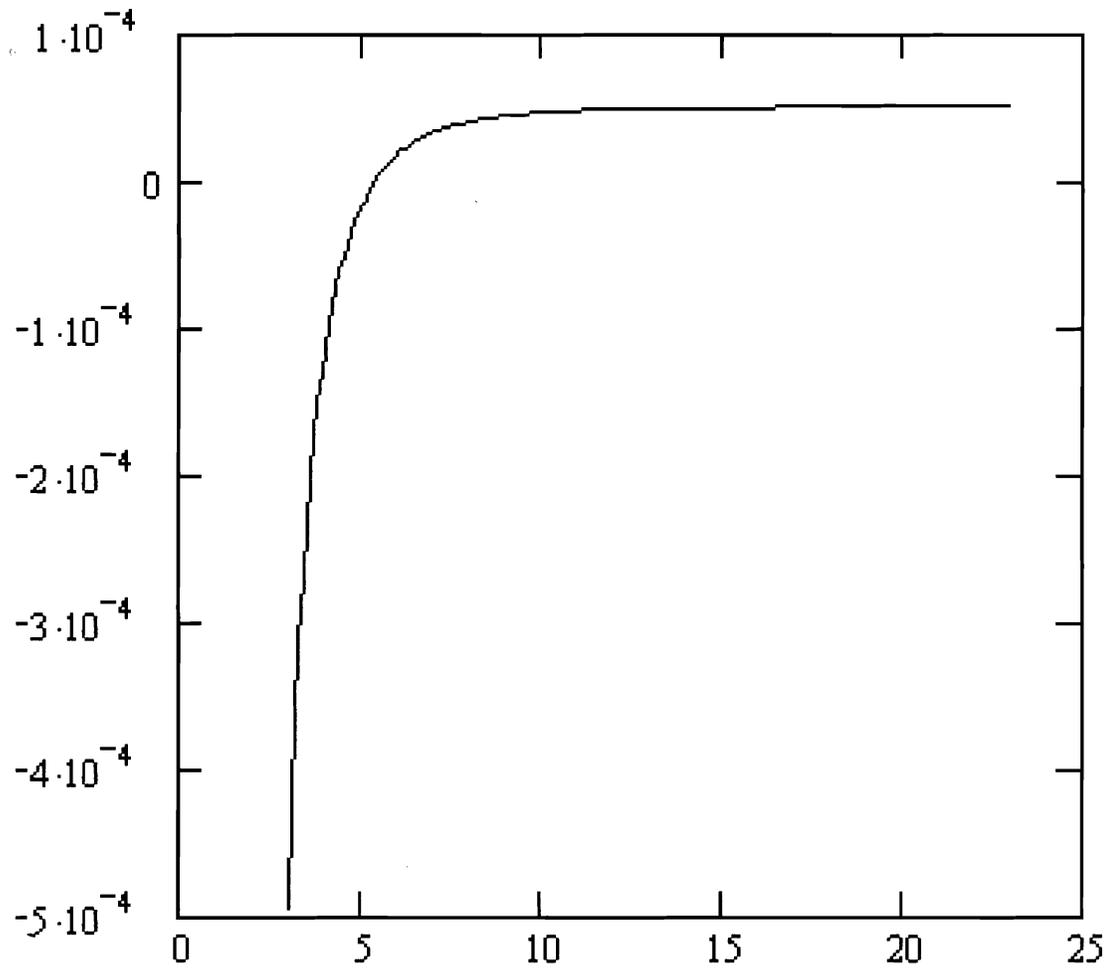}
  \caption{The Casimir force versus $\frac{R}{L}$
  between two parallel plates in the spacetime with one compactified extra dimension.}
\end{figure}

\newpage
\begin{figure}
\setlength{\belowcaptionskip}{10pt} \centering
  \includegraphics[width=15cm]{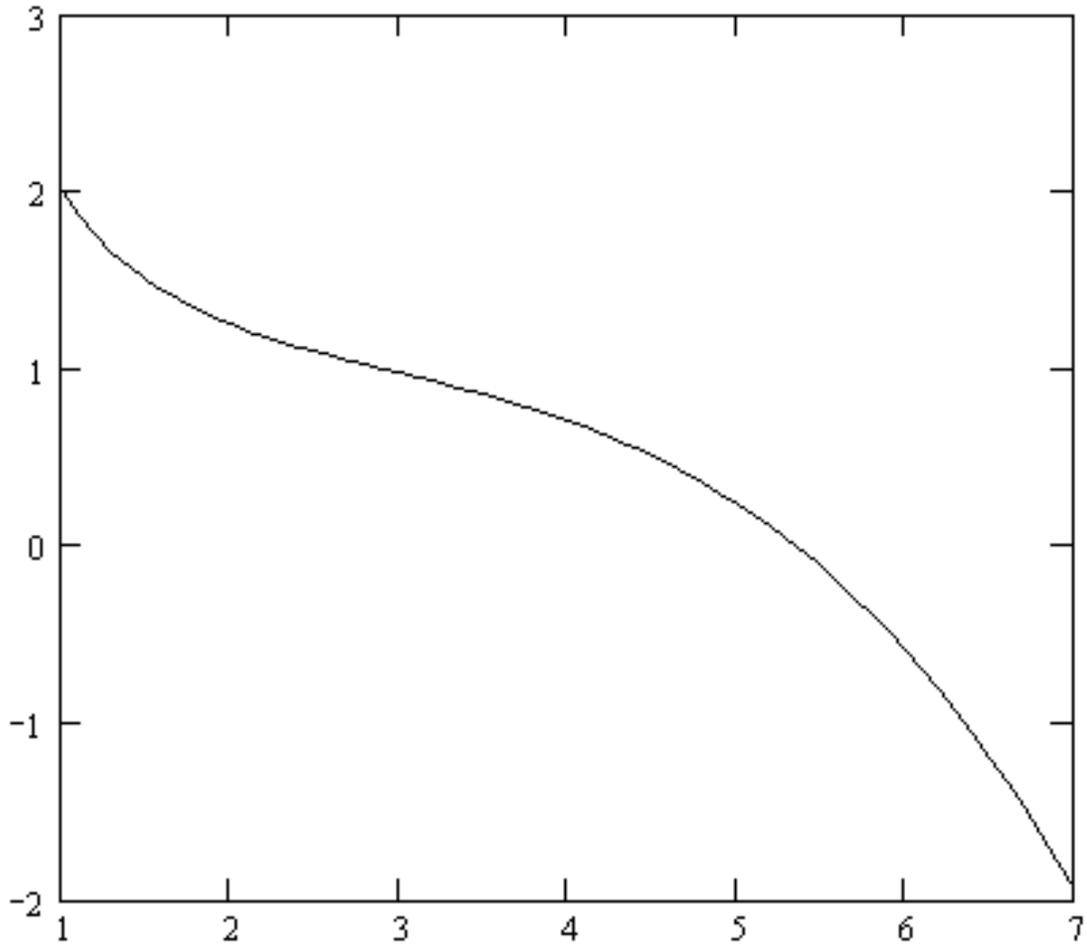}
  \caption{The ratio of the Casimir force (with one extra dimension) to the standard ones (without extra dimension)
  versus $\frac{R}{L}$ between two parallel plates.}
\end{figure}

\end{document}